\documentclass[10pt,a4paper]{article}
\usepackage[utf8]{inputenc}
\usepackage{amsmath}
\usepackage{amsfonts}
\usepackage{amssymb}
\usepackage{subfigure}
\usepackage{multirow}
\usepackage{graphicx}
\usepackage[left=2cm,right=2cm,top=2cm,bottom=2cm]{geometry}
\begin{document}
\title{Antara: An Interactive 3D Volume Rendering and Visualization Framework}

\author{Pratik Kalshetti \and Parag Rahangdale \and Dinesh Jangra \and Manas Bundele \and Chiranjoy Chattopadhyay\\
Indian Institute of Technology Jodhpur}

\maketitle

\begin{abstract}
The goal of 3D visualization is to provide the user with an intuitive interface which enables him to explore the 3D data in an interactive manner. The aim of the exploration is to identify and analyse anomalies or to give proof of the non-anomaly of the visualized organic structures. For 3D Medical Data, Magnetic Resonance Images (MRI) have been used. To create the 3D model, we used the Direct Volume Rendering technique. In the input 3D data, we have $x, y$ and $z$ coordinates and an intensity value for each voxel. The 3D data is used by Volume Ray Casting to compute 2D projections from 3D volumetric data sets. In ray casting, a ray of light is made to pass through the volume data. The interaction of each voxel with this ray is used to assign RGB and alpha values for every voxel in the volume. As a result, we are able to generate the 3D model of the region of interest using the 3D data. The 3D model is interactive, thus enabling us to visualize the different layers of the 3D volume by adjusting the transfer function.
\end{abstract}


\section{Introduction}
In medical field, visualizing the internal structure of body is very important for proper medical diagnosis \cite{post2012data}. To aid doctors in visualizing the internal body parts, scanning methods like MRI (Magnetic Resonance Imaging), X-ray and CT (Computed Tomography) scan are used. Output of such methods are generally one or more grayscale 2D images\cite{34710}. It is difficult to visualize the internal structure with the help of these 2D images.

\begin{figure}
\centering
\includegraphics[scale=0.2]{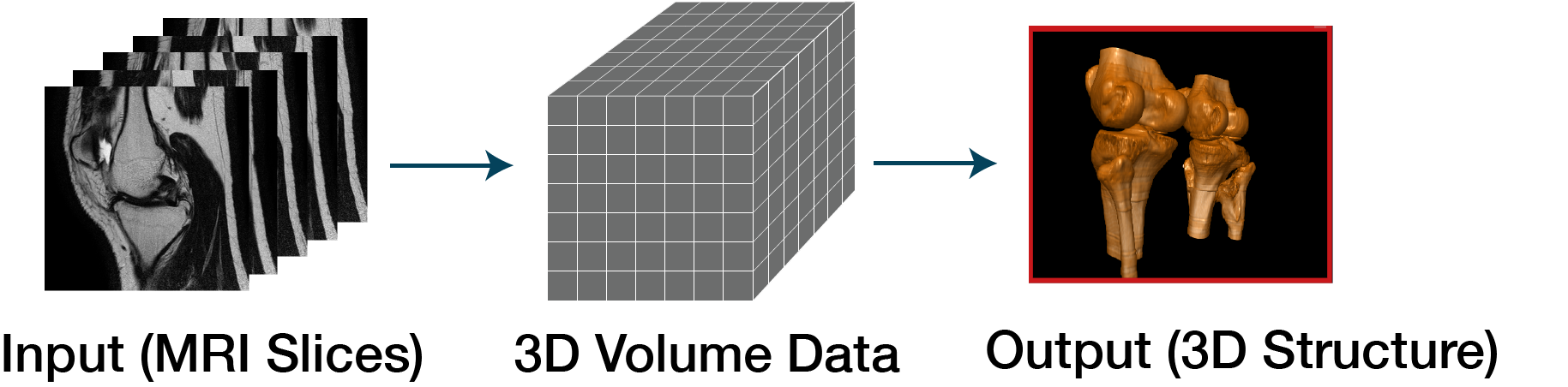}
\caption{Extracting 3D bone from MRI of knee.}
\label{figure1}
\end{figure}

To solve this problem, we propose an interactive 3D volume rendering and visualization framework Antara (Sanskrit: Interior) from the data obtained using MRI scan as shown in Fig. \ref{figure1}. This has been achieved using Volume Rendering \cite{rendering}. Volume rendering is an important graphics and visualization technique. A rendered volume can be used for displaying not only surfaces of a model but also the intricate details contained within. In order to achieve our goal, we chose Direct Volume Rendering (DVR) technique for 3D visualization. It helps to generate an interactive 3D model which can be altered to show surfaces having different intensities\cite{hohne19903d}. DVR methods generate images of a 3D volumetric data set without explicitly extracting geometric surfaces from the data. These techniques use an optical model to map data values to optical properties, such as color and opacity. During rendering, these optical properties are accumulated along each viewing ray to form an image of the data. This helps in generating accurate images of the part being visualized \cite{raman2013mdct}.

\section{Literature Survey}
Most of the fundamental volume visualization algorithms can be divided into two categories: DVR algorithms and Surface Fitting (SF) algorithms. SF algorithms fit (usually planar) surface primitives such as polygons or patches to constant-value contour surfaces in volumetric data-sets. The user begins by choosing a particular threshold value and the geometric primitives are then automatically fit to the high contrast contours in the volume that match the threshold. Cells whose corner values are all above the chosen threshold or all below the threshold are removed and thus have no effect on the final image. SF algorithms constitute few of the algorithms like opaque cubes, marching cubes, dividing cubes and marching tetrahedra\cite{okuyan2014direct}. SF methods are typically faster than DVR methods since SF methods only traverse the volume once to extract the surfaces. After extracting the surfaces, rendering hardware and other rendering methods can be used to quickly render the surface primitives each time the user changes a viewing or lighting parameter. Changing the SF threshold value is time consuming because it requires that all of the cells be revisited to extract a new set of surface primitives. Also, SF methods suffer from several problems such as occasional false positive and negative surface pieces and incorrect handling of small features and branches in the data. Also, it only displays surfaces which meet a threshold density and will only show the surface that is located closest to the imaginary viewer.

DVR methods perform mapping of elements directly into the screen without using geometric primitives as an intermediate representation. DVR algorithms include approaches such as ray-casting, integration methods, splatting, and V-buffer rendering. In DVR method, transparency and colors are used to allow a better representation of the volume to be depicted in a single image. DVR retains all of the volume data during rendering, and thus all the acquired voxels may contribute to each rendered image. Spatial gradients is computed prior to or during rendering, and these gradients, as well as the relative contributions of each voxel, is adjusted to produce images of surfaces. Thus volume rendering can produce images of surfaces. Therefore, if volume rendering of medical data sets is inexpensive and fast enough, surface rendering will no longer be utilized.
\begin{figure}[!b]
\centering
\includegraphics[scale=0.5]{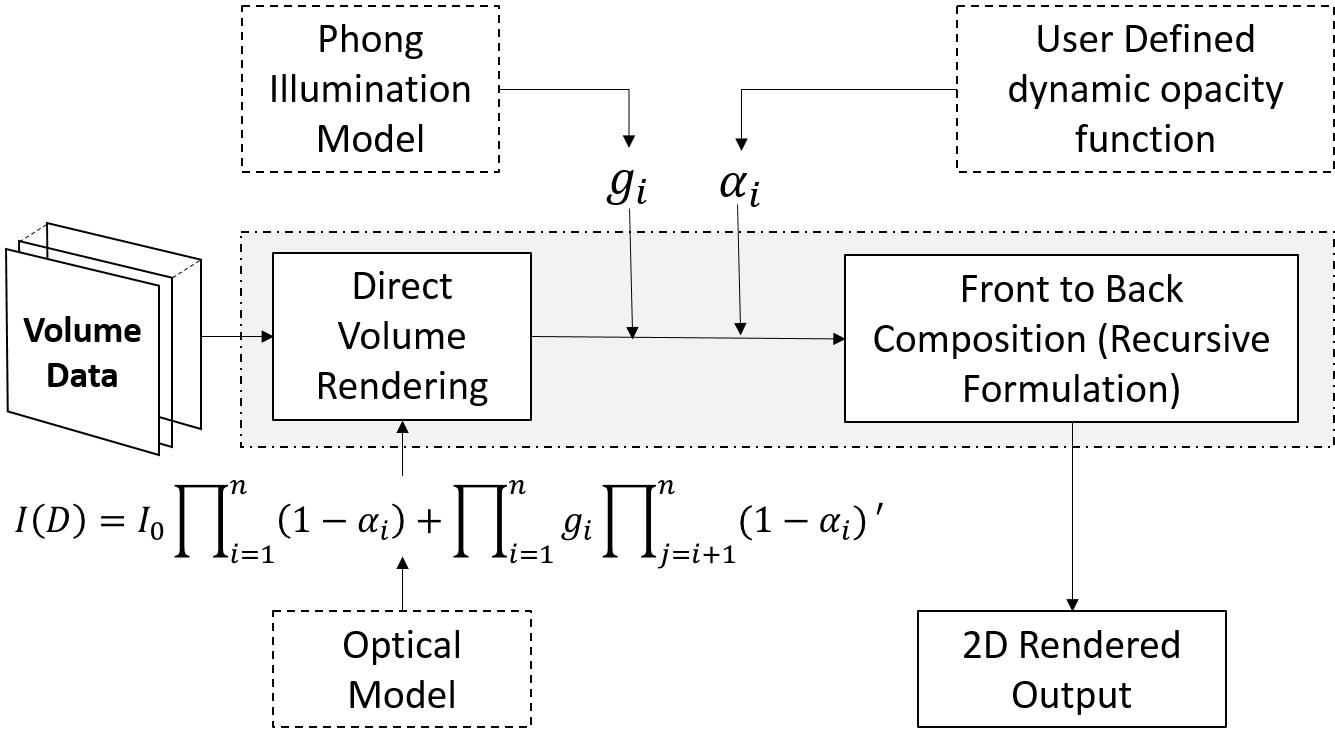}
\caption{Overall framework of Antara.}
\label{figure2}
\end{figure}

Ray-casting\cite{ray} is the popular technique of volume visualization for the production of high quality images. It conducts an image-order traversal of the image plane pixels, finding a color and opacity for each. The color tuple and opacity values are added to the pixel using a weighting formula and a step is taken along the ray to the next re-sample point. In ray casting, rays continue in a straight line until the opacity encountered by the ray sums to unity or the ray exits the rear of the volume. No shadows or reflections are generated in ray-casting. Because the magnitude of the gradient is usually a good indication of the strength of an iso-surface within a cell, the result is a large color contribution to the pixel when the ray encounters a new substance in the volume\cite{soler2014real}. Ray-casting is CPU-intensive, but the images show the entire data-set, not just a collection of thin surfaces as in SF. 3D models were created using Ray casting algorithm \cite{adeshina2012hardware} which is a volume rendering technique. Volume rendering techniques are computationally intensive, whereas the ray tracing algorithm \cite{hachisuka2008multidimensional,levoy1990efficient, parker2005interactive}, is faster in terms of running time and avoiding redundant computations. 

The important objectives in visualizing medical data is high quality rendering and provide real time interaction. This paper highlights the use of ray tracing for volume rendering in the medical domain for getting an insight into the interior stucture of the organs of the body using DVR, thus producing high quality rendered image focusing even minute anomalies in the bones and tissues. Also the real time interaction is achieved with the recusive formulation of the rendering equation. Another novel highlight of the paper is the metric used for analysing the result of the volume rendering. The idea is an extension of the image segmentation metric used in 2D images, thus opening gates to use such metrics for 3D data.

\section{System Overview}
The overall approach comprises of the modules explained in Fig. \ref{figure2}. Representing a surface contained within a volumetric data set using geometric primitives can be useful in many applications, however, there are several main drawbacks to this approach \cite{Gordillo20131426}. First, geometric primitives can only approximate surfaces contained within the original data. Adequate approximations may require an excessive amount of geometric primitives. Therefore, a trade-off must be made between accuracy and space requirements. Second, since only a surface representation is used, much of the information contained within the data is lost during the rendering process \cite{Elvins:1992:SAV:142413.142427}. For example, in CT scanned data useful information is contained not only on the surfaces, but within the data as well. Also, amorphous phenomena, such as clouds, fog, and fire cannot be adequately represented using surfaces, and therefore must have a volumetric representation, and must be displayed using volume rendering techniques.

\subsection{Volumetric Function Interpolation}
The volume grid $V$ only defines the value of some measured property $f(x,y,z)$ at discrete locations in space. If one requires the value of $f(x,y,z)$ at an off-grid location $(x,y,z)$, a process called interpolation must be employed to estimate the unknown value from the known grid samples $V(x,y,z)$. We  have used trilinear interpolation in this case. It can be written as 3 stages of 7 linear interpolations, since the filter function is separable in higher dimensions. The first 4 linear interpolations are along $x$:
\begin{equation} \label{eq1}
f(u,v_0,1,w_0,1) = (1 - u)(f(0,v_0,1,w_0,1)) + uf(1,v_0,1,w_0,1)
\end{equation}
Using these results, 2 linear interpolations along $y$ follow:
\begin{equation} \label{eq2}
f(u,v,w_0,1) = (1 - v)(f(u,0,w_0,1)) + uf(u,1,w_0,1)
\end{equation}
and, interpolation along $z$ yields the result:
\begin{equation} \label{eq3}
f(x,y,z) = f(u,v,w) = (1 - w)(f(u,v,0)) + wf(u,v,1)
\end{equation}

Here the $u,v,w$ are the distances (cell of size of 13 (empirically determined)) of the sample at $(x,y,z)$ from the lower left rear voxel in the cell containing the sample point. 

\begin{figure}[t]
\centering
\includegraphics[scale=0.09]{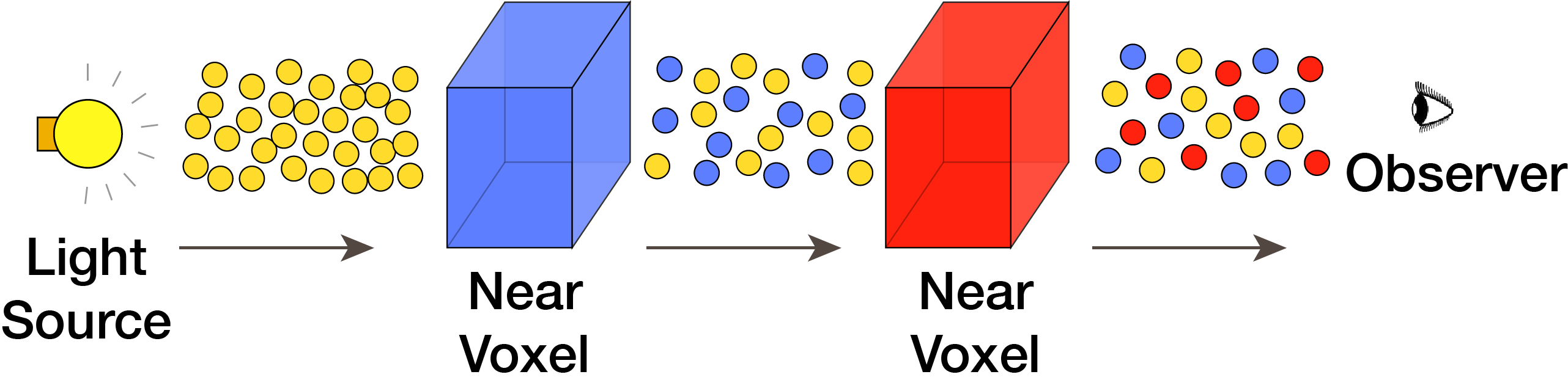}
\caption{Optical model used for characterizing the optic properties of the volume.}
\label{figure3}
\end{figure}

Optical models for direct volume rendering view the volume as a cloud of particles. Light from a source can either be scattered or absorbed by particles. The particles in an actual cloud occlude incoming light, as well as add their own glow as shown in Fig. \ref{figure3}. 

\subsection{Direct Volume Rendering}
Volume rendering or DVR is the process of creating a 2D image directly from 3D volumetric data, hence it is often called direct volume rendering\cite{volumerendering}. We have used an image-order technique. It includes the following operations:
\begin{enumerate}
  \item Cast rays from the image pixels, sampling the grid at discrete locations along their paths.
  \item Obtain the samples via interpolation.
  \item The interpolated samples are further processed to simulate the light transport within a volumetric medium.
\end{enumerate}
The fundamental element in DVR is the volume rendering equation. It is obtained by using an absorption and emission model for particles. The equation is formulated as partial differential equation which is then solved by Ordinary Differential Equation (ODE) solver. Using the Riemann approximation for the integral present in the solution of the partial differential equation, we get the discretized version of the same. The discrete version of the rendering equation \cite{rendering} at the distance $D$ of the image plane is given by:
\begin{equation} \label{eq4}
I(D) = I_0\prod_{i=1}^n(1-\alpha_i) + \sum_{i=1}^{n}g_i\times\prod_{j=i+1}^n(1-\alpha_i)
\end{equation}
where $I_0$ is the background light, $\alpha_i$ and $g_i$ is the opacity and color value associated the sampled volume data points assuming $n$ sampled data points.

Note that the first term in Eq. \ref{eq4} captures absorption whereas the latter one focuses on the emission. The above equation requires redundant computation for each of the data points which can be resolved by using recursive implementation called the Front-to-Back Compositing which is defined in Eq. \ref{eq5} and Eq. \ref{eq6}. 
\begin{equation} \label{eq5}
C_{out} = C_{in} + C(x)\alpha(x)\times(1 - \alpha_{in})
\end{equation}
This formulation makes it possible to implement the approach and perform the process in real time.
\begin{equation} \label{eq6}
\alpha_{out} = \alpha_{in} + \alpha(x)\times(1 - \alpha_{in})
\end{equation}
Here, $C_{in}$ and $C_{out}$ denote the color value before and after any iteration and similarly $\alpha_{in}$ and $\alpha_{out}$ denote the opacity value before and after any iteration.

\subsection{Raycasting}

In nature a light source is always required to visualize any object. Raycasting algorithm is used to determine the color and opacity associated with each voxel present in volume data due to the light source\cite{5307637}. It is achieved by appling following operation on each voxel \cite{ray}:
\begin{enumerate}
  \item Cast a ray into the volume
  \item Linearly interpolate data values from voxel corners
  \item Convert data values to optical properties viz. color and opacity
  \item Composite the optical properties
  \item Return the final color to assign to the image plane pixel
\end{enumerate}

\begin{figure}
\centering
\includegraphics[scale=0.13]{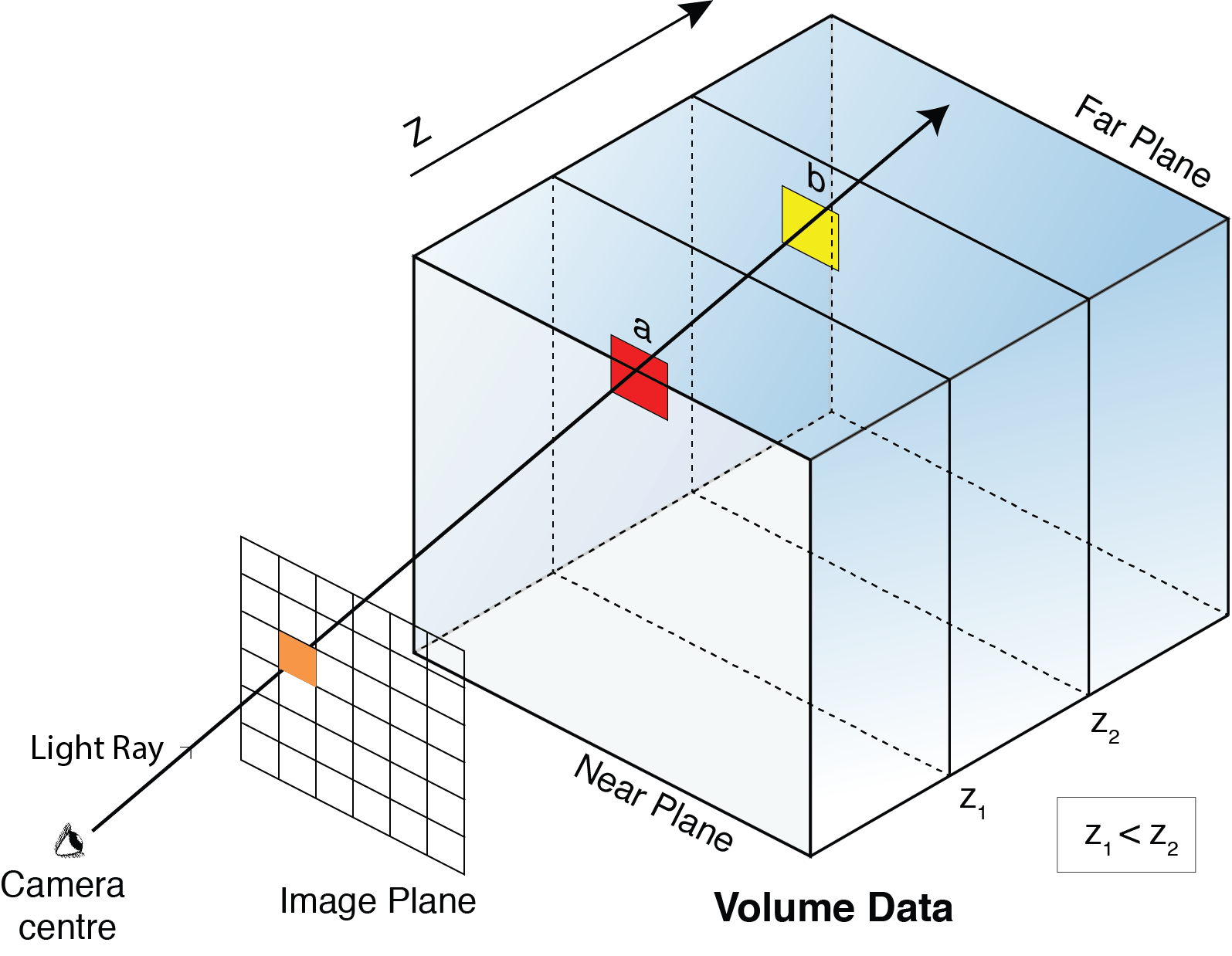}
\caption{Illustration of the ray casting technique.}
\label{figure4}
\end{figure}

The above steps are shown in Fig. \ref{figure4}.

\subsection{Shading}
We have explored various shading models and found Phong Illumination Model (see Eq. \ref{eq8}) best suitable for the purpose of obtaining the color value $g$ associated with each sampled data value. This is further emphasised in Sec. \ref{sec:shading}. Here, the normal is computed using central difference method to find the gradient.

\begin{equation} \label{eq8}
g_i = C(x_i)\times I_a + C(x_i)\times I_d \times (N.L) + C(x_i)\times I_s \times (R.V)^n
\end{equation}
where $C(x_i)$ is the color of the sample $i$; $I_a$, $I_d$, $I_s$ are the light's ambient, diffuse, and specular components, $N$ is the normal at sample $i$, $V$ is the vector from sample point to eye, $L$ is the light vector from the sample to light source, $R$ is the reflection vector and $n$ is the shininess. Now once obtained the color value, the next task is to obtain the opacity value which can be obtained from the opacity function.

\begin{figure}[!b]
\centering
\includegraphics[scale=0.12]{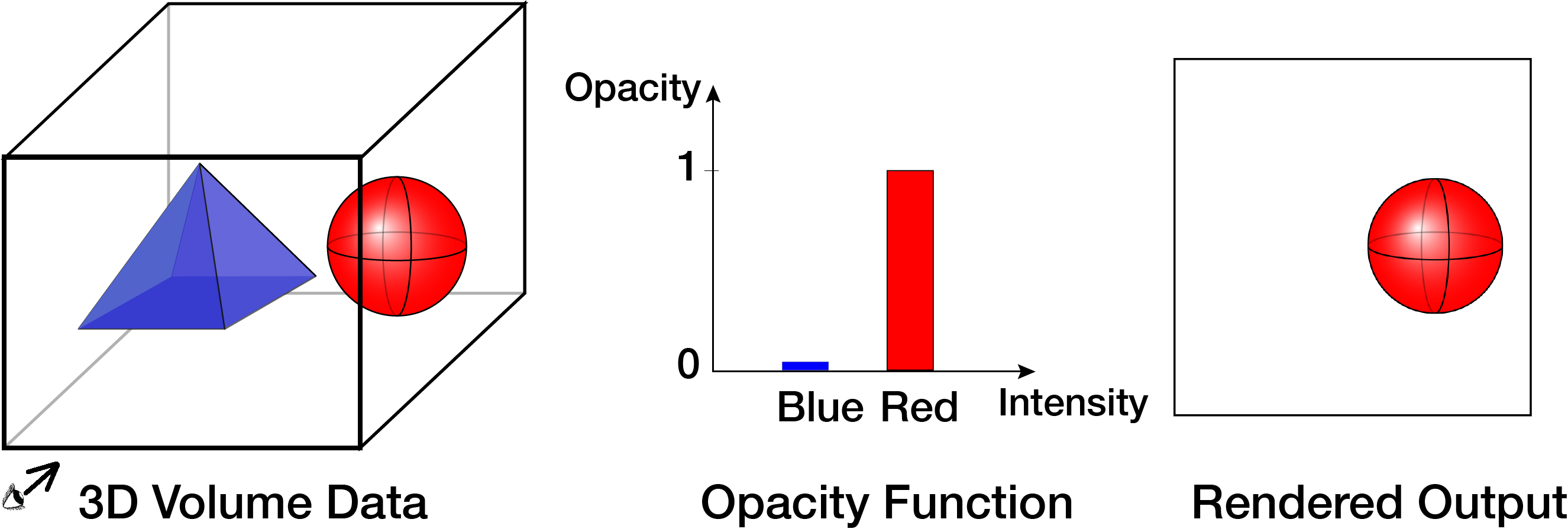}
\caption{Effect of varying opacity on rendered objects.}
\label{figure6}
\end{figure}

\subsection{Opacity  Function}
This stage computes the opacity value of different data points within the sampled volume. The user provides the opacity value with the help of a transfer function defined as $f(x)= \alpha(x)  \forall  x\in[1,...,n]$ where $\alpha(x)$ is the opacity value for all the data points with intensity value of $x$. The rationale behind providing this user interaction is to clearly view the region of interest by setting its opacity to a larger value. Finally, the compositing values obtained, thus forms the pixel values for the image plane which projects the 3D volume data onto a 2D plane. 
\begin{figure*}
\centering
\includegraphics[scale=0.2]{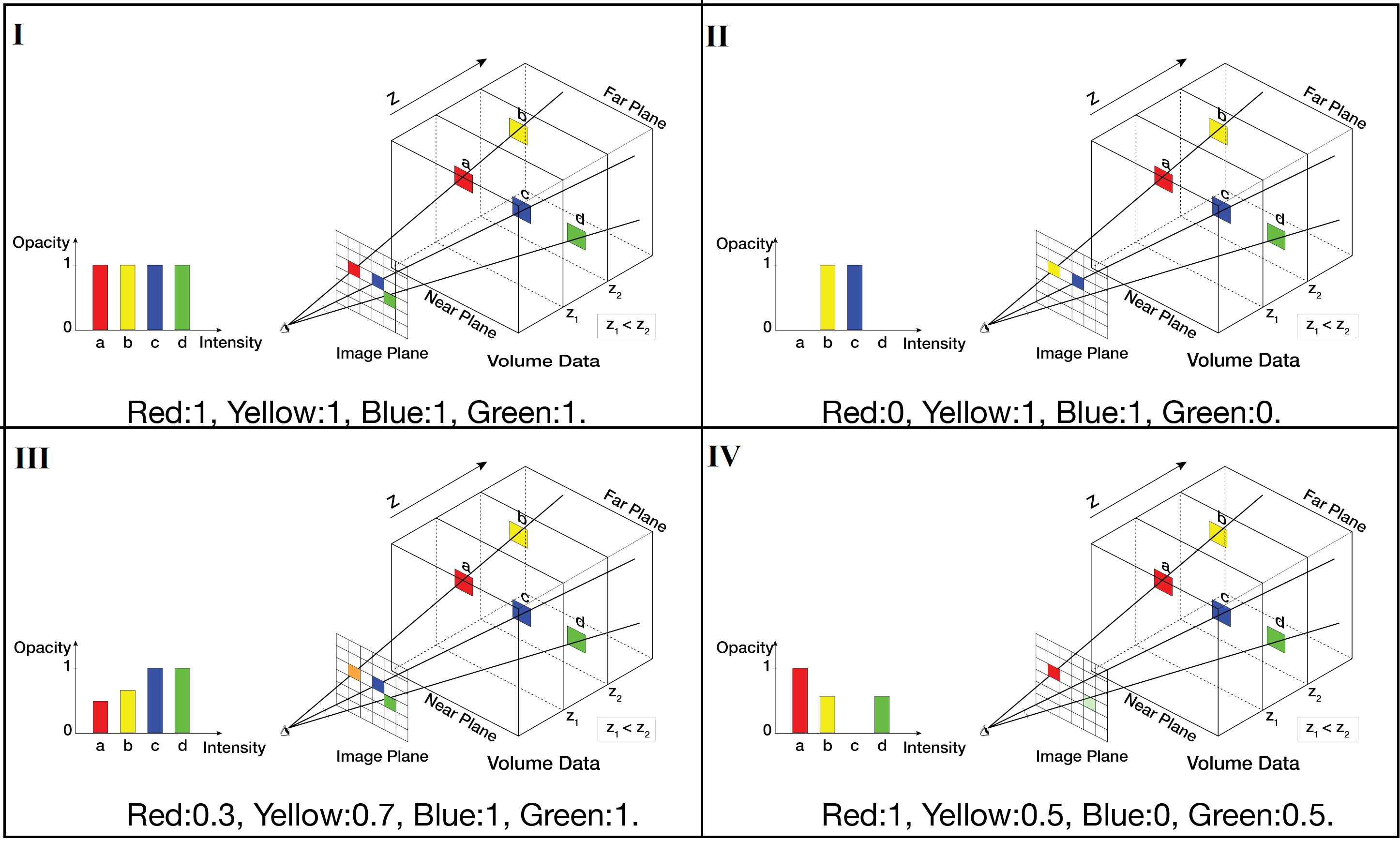}
\caption{Illustration of Antara's rendering process. (I)-(IV): effect of varying opacity values of different voxels.}
\label{figure7}
\end{figure*}
An illustration of this stage is shown in Fig. \ref{figure6}. Consider two objects in the volume data. The pyramid is blue in color and the sphere is red. When viewed from the bold face of the cuboid as shown, only the sphere is rendered on the image plane since the opacity value for red intensity data points is set to a high value whereas that for blue data points is set to a low value.

\subsection{Rendering Algorithm}
The overall idea is depicted in Fig. \ref{figure7}. Consider a volume data consisting of four data points with intensity values being red(a), yellow(b), blue(c) and green(d). The blue and red points lie at same depth($z_1$) while green and yellow lie at a different depth($z_2$) and the former depth being nearer to the screen. The red and the yellow data point lie on the same ray along the camera to a pixel on the image plane. The opacity function shows the opacity values for each of the intensity values present in the volume data. Since the red and yellow data points are projected on the same pixel on the image, the final value at the pixel is a weighted combination of these two values where the weights are determined by the opacity function which is modified by the user.

The steps have been illustrated below:
\begin{enumerate}
\item For full opacity $(value=1)$, intensities $a$, $c$ and $d$ are visible on the image plane. Intensity $b$ has no effect because it is behind $a$ (see Fig. \ref{figure7}(I)).
\item Zero opacity value for $a$ and $d$ hides them completely in the image plane and intensity $b$ (see Fig. \ref{figure7}(II)).
\item Setting $0<a,b<1$, a mixture of both the intensities is visible on the image plane (see Fig. \ref{figure7}(III)).
\item When $a=1$, varying $b$ has no significance as it is obstructed completely by $a$ (see Fig. \ref{figure7}(IV)).
\end{enumerate}

\section{Experimental Results}

\subsection{Dataset}
The dataset used for evaluation of the proposed approach is the volume data obtained via MRI and CT of 5 human organs (brain, foot, knee, tooth and heart). The ground truth is created by manually segmenting the bone region from each of the slices(images) from the scan and then validated by a medical expert. Also we have scans of an engine and a car for proving the proposed algorithm's generalization capability in other domains. 

\subsection{Qualitative Analysis}
\begin{figure}[!h]
\centering
\includegraphics[scale=0.2]{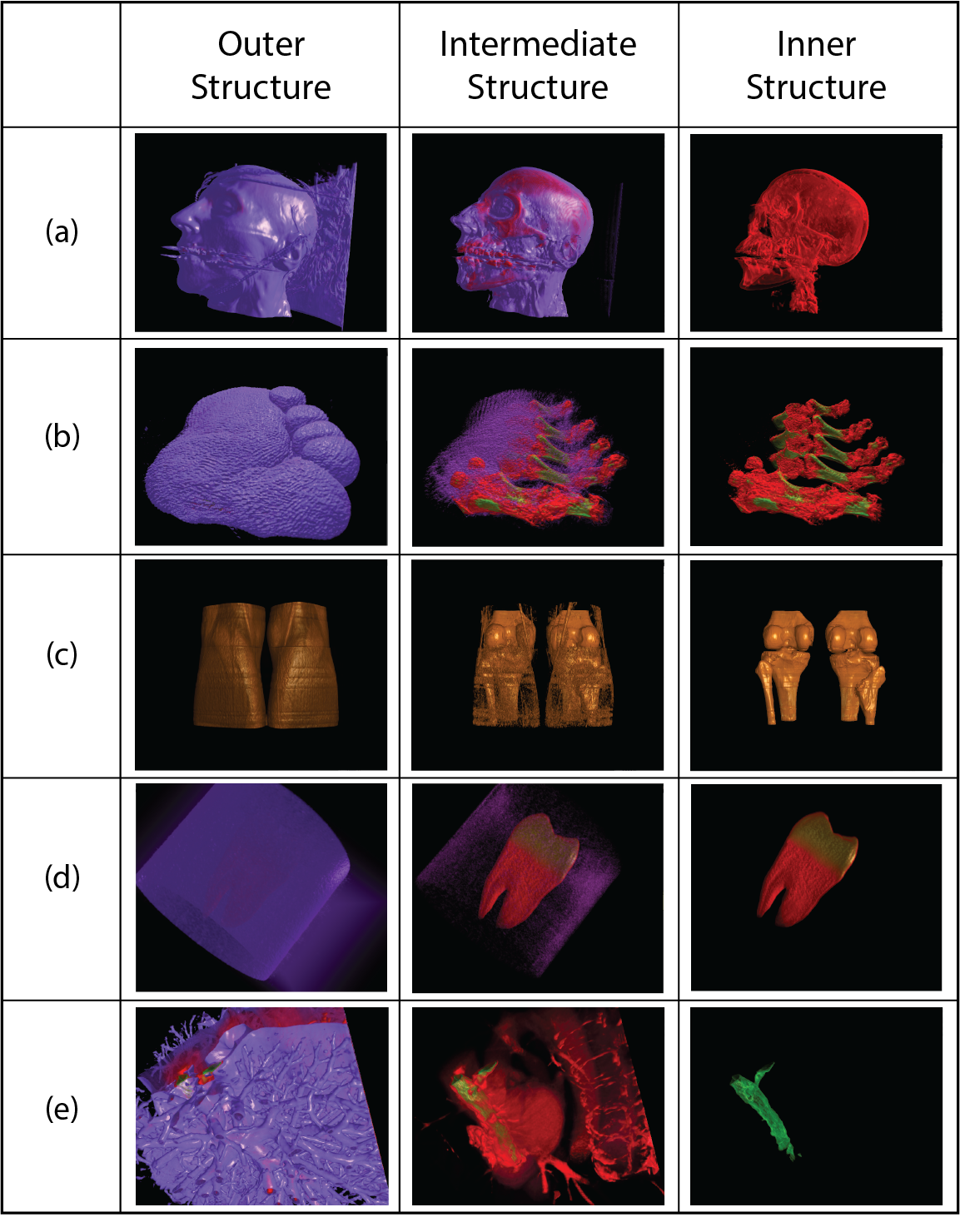}
\caption{Result on different datasets: (a) Brain CT, (b) Foot CT, (c) Knee MRI, (d) Tooth CT, (e) Heart CT.}
\label{figure8}
\end{figure}
Antara's performance on different datasets is shown in Fig. \ref{figure8}. The colors in the figure are chosen for better visualization. Changing the opacity function for different intensities varies the rendered object, thus giving an insight into the different structures as per the material characteristics. In Fig. \ref{figure8}, the outer structure is obtained by giving high opacity values to them and gradually decreasing this value for the outer structures helps in visualizing the inner structure. The obtained results can this aid in understanding the hidden structure and thus find flaws in it. This will help in better surgery as per the patient needs. The proposed method achieves the objective of high accuracy required in the visualization of medical images, which can render non geometrical shapes with precision. The removal of tissues yields exact visualization of bones, which is possible due to their relative intensity difference.

\subsection{Generalization Capability}
\begin{figure}[h]
\centering
\includegraphics[scale=0.23]{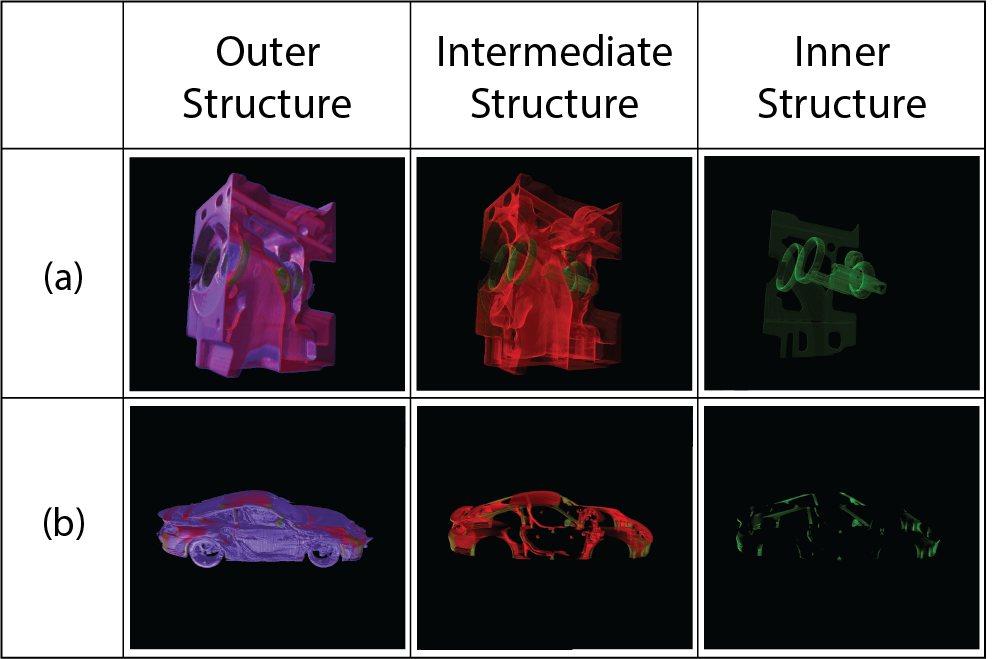}
\caption{Result on automobile volume data.}
\label{figure12}
\end{figure}

Antara not only works well with medical data but also works on data from other domains. In Fig. \ref{figure12} data from automobile sector is worked upon by the proposed algorithm and shows high quality result giving insight into internal structures. This data includes a porche car scan and an engine volume data. This aids in detecting flaws in the internal structure of the machine. This shows that Antara has far reaching effects on different disciplines like city architecture plans, electrical systems and industrial machines.

\subsection{Quantitative Analysis}

The qualitative analysis clearly depicted the efficacy of Antara in the previous section. Quantitative analysis, in terms of accuracy of the proposed algorithm, provides a better analysis of the technique. To evaluate the performance of our segmentation approach, we have used the Dice similarity coefficient index as a measure to quantify the accuracy of the obtained results, where the ground truth has been generated manually and later validated by the doctor.

The adopted technique is a novel method for analysing the volume data. The genesis of this technique can be traced back from the fact that dice coefficient is a technique widely used in validation of segmentation of medical images. Extending this idea, and with the thought that the volume data in MRI is a formed by slice wise images of a particular organ, the rendered image can then be considered as a result obtained by segmenting the corresponding slice from input volume data(MRI slices). The Dice coefficient is given by:

\begin{equation} \label{eq9}
d = 2 * \dfrac{|R_{seg} \cap R_{gt}|}{|R_{seg}| + |R_{gt}|}  
\end{equation}

where $R_{seg}$ is the segmented result of Antara and $R_{gt}$ is the manually segmented ground truth.\\ 

\begin{figure}[t]
\centering
\includegraphics[scale=0.63]{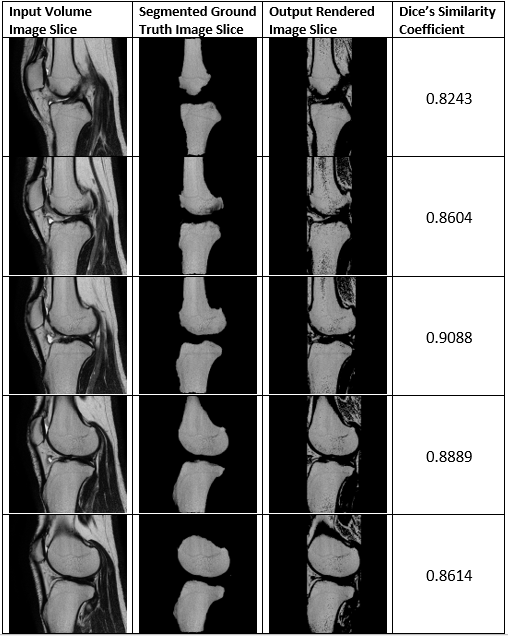}
\caption{Quantitative Evaluation.}
\label{figure9}
\end{figure}

As shown in Fig. \ref{figure9}, the dice coefficient of a slice obtained from the rendered volume when compared with the ground truth slice containing only the bone results into a value close to 1 for almost all the slices. Thus, this technique provides higher accuracy results which are essential in the medical domain. The average dice similarity coefficient index obtained over the entire data set used, was found to be $0.8176$. 

This value clearly proves the efficiency of the results generated by the proposed approach and hence making it suitable for using in applications where high precision is required.
This principle of measuring the result of a 3D data using the methods available for 2D data is in itself an innovation in the analysis phase in the domain of computer graphics.

\subsection{Effect of shading and shininess}
\label{sec:shading}
\begin{figure}
\centering
\begin{tabular}{cc}
\includegraphics[scale=0.2]{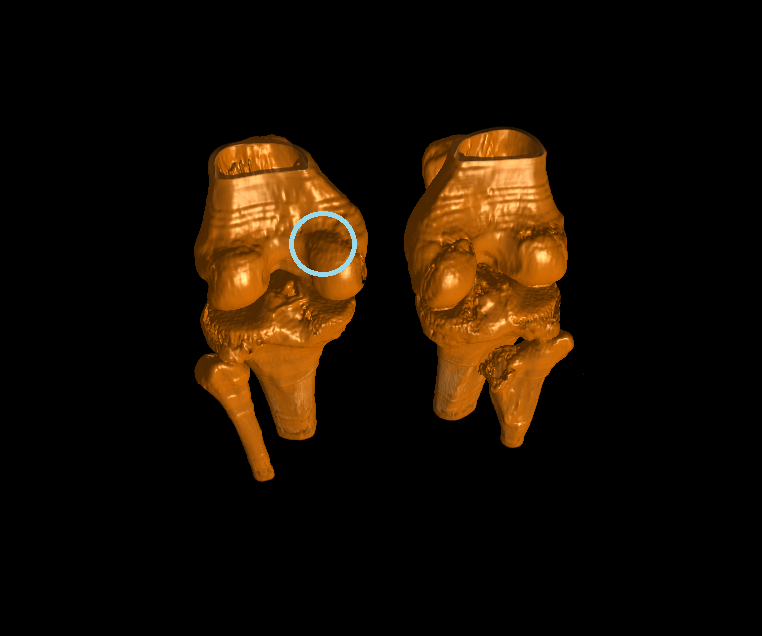}&
\includegraphics[scale=0.2]{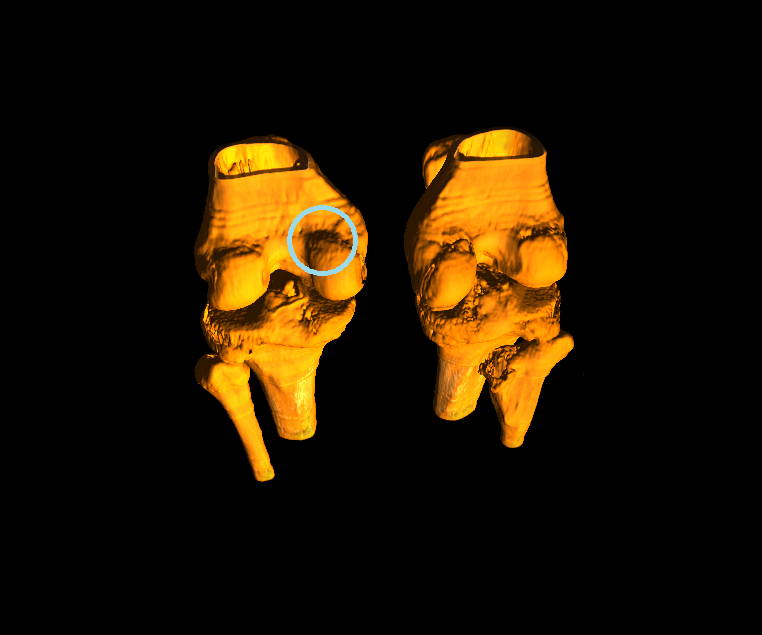}\\
Phong Model&Cel Model
\end{tabular}
\caption{Comparison between Phong and Cel Model.}
\label{figureshine}
\end{figure}

Shading plays an important role in visualizing a scene. Human vision uses shading as a cue to form position and depth. Total handling of light is very expensive and shading model gives a good approximation of what would really happen much less expensively. The two less expensive techniques (i.e. real time interactive) widely used are Phong model and the Cel shading model. The result obtained by Phong model helps better visualizing the data points that are occluded from the scanning direction as compared to the Cel model. As shown in Fig. \ref{figureshine} the encircled region is not highlighted clearly in the Cel Model, while it is shaded properly to visualize it appropriately.

\begin{figure}[t]
\centering
\includegraphics[scale=0.13]{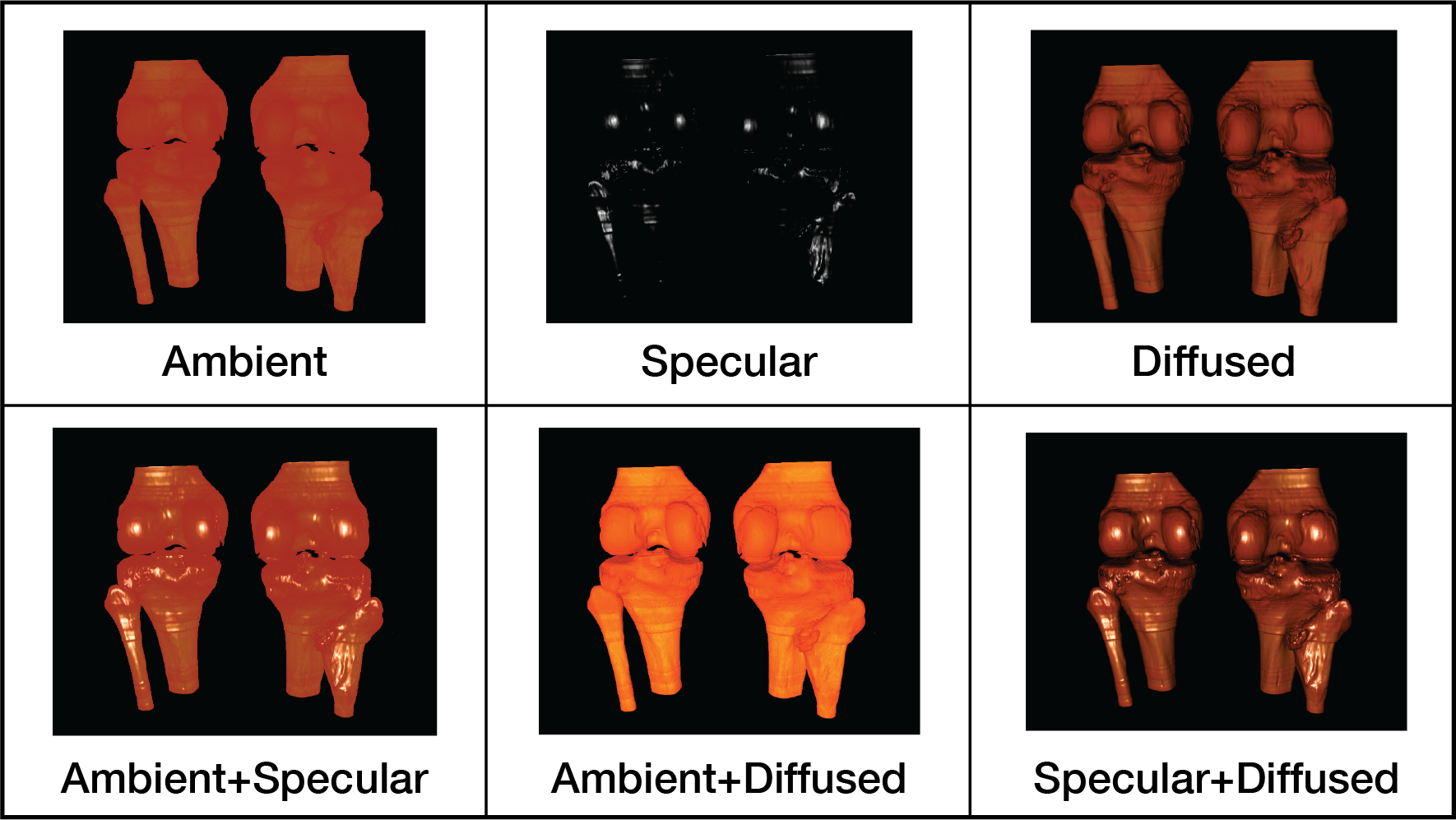}
\caption{Effect of shading on the rendered object.}
\label{figure10}
\end{figure}

The importance of incorporating all the three lights viz. ambient, diffused and specular is clear from Fig. \ref{figure10}, where the rendered image by using various combination of these lights is shown. All these images proves that none of these lighting models when used alone or pairwise provide reallistic results, thus making it a necessity to use all the three lights together to provide high quality realistic result.

\begin{figure}[t]
\centering
\includegraphics[scale=0.16]{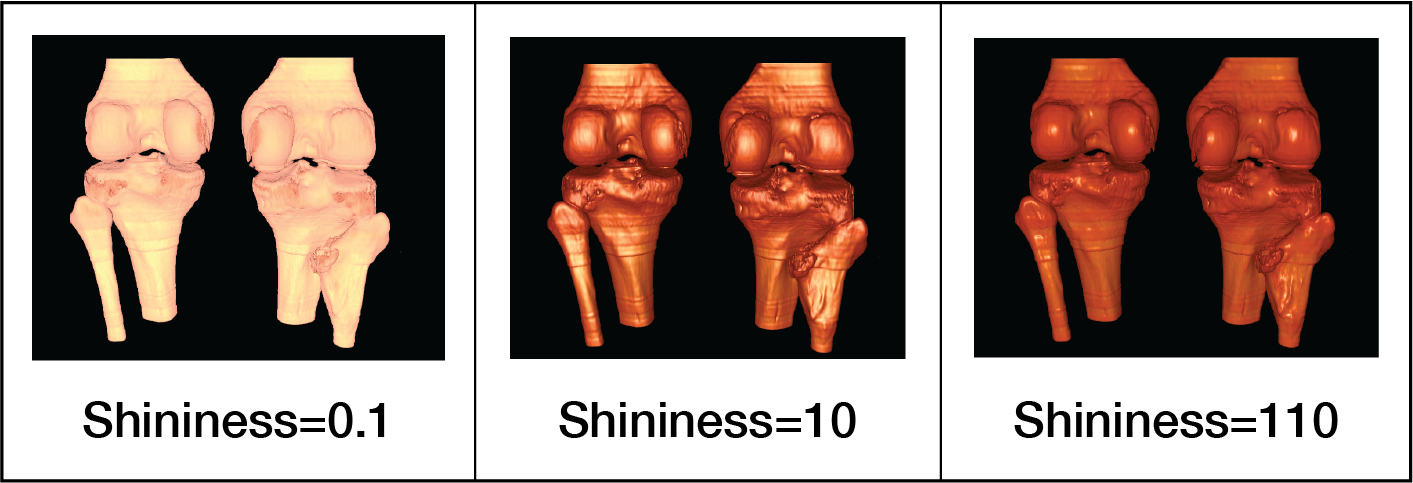}
\caption{Effect of varying shininess values on rendering.}
\label{figure11}
\end{figure}

Shininess is the coefficient of specular reflectivity associated with the surface of an object being rendered. Appropriate coefficient needs to be chosen as per the application. All the results shown throughout have been generated with the value of 60 for the coefficient. The variation of shininess has been shown visually in Fig. \ref{figure11}. 

\begin{figure}[!b]
\centering
\begin{tabular}{ccc}
\includegraphics[scale=0.13]{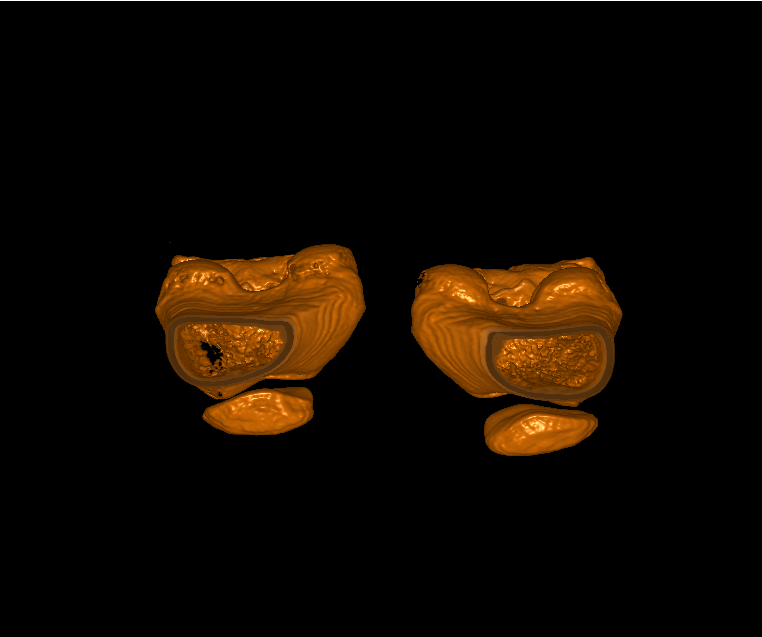}&
\includegraphics[scale=0.13]{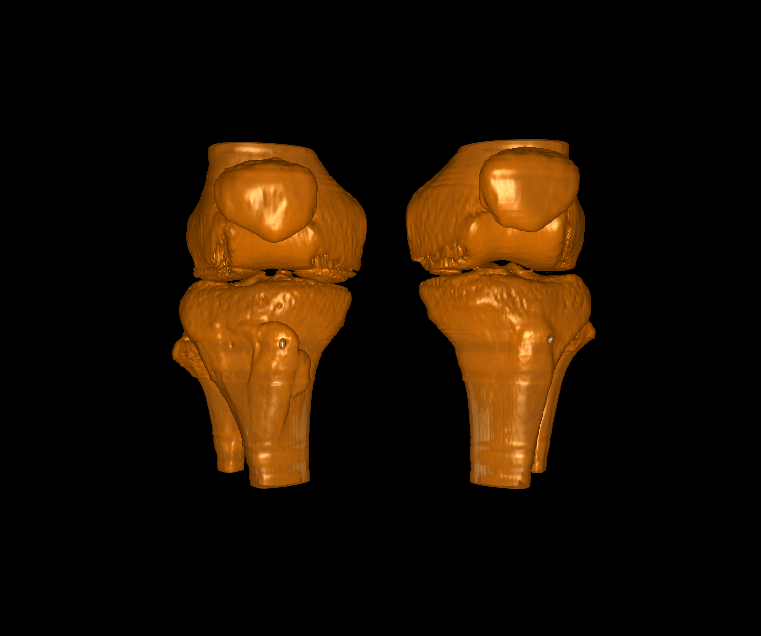}&
\includegraphics[scale=0.13]{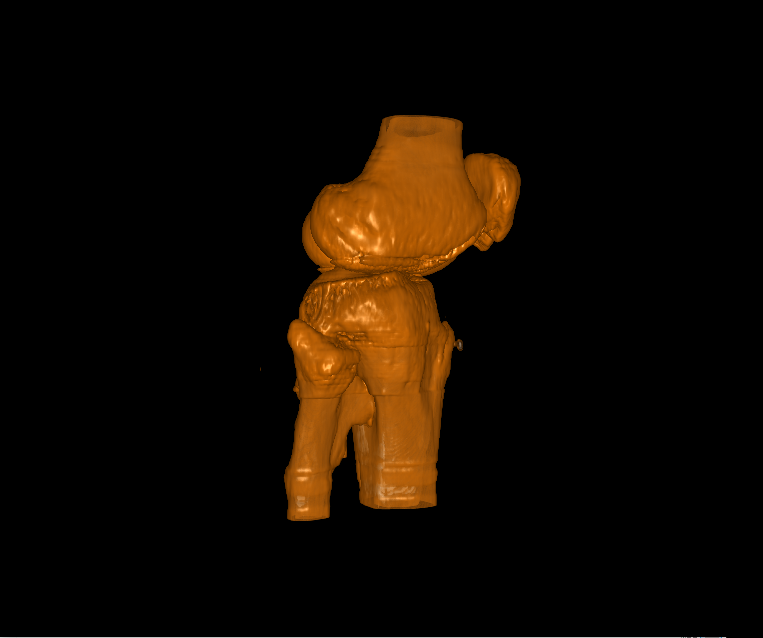}\\
Axial&Coronal&Sagittal
\end{tabular}
\caption{Different Views of Knee.}
\label{figure13}
\end{figure}

The rendered image in Fig. \ref{figure13} with all the lighting components included and a value of $60$ for the shininess coefficient gives an accurate result obtained in real time. The results are generated from the MRI scan of the knee when viewed from different orthogonal planes. The color is chosen for better visualization. The recommended value for shininess is 60 using the Phong Shading model with all the three light components for the medical dataset. 

\section{Conclusions}
Antara works on the basis of differences in the intensities of the components of the 3D data. It is a multi-purpose platform for the analysis of various types of patient data. E.g., analysis of bones as well as tissues, detection of tumors, etc. Antara also works for non-medical domain also (e.g. automobiles). There is a dearth of such low cost, open-source software available in the market for analyzing 3D models of medical images. Such softwares can be used for image guided surgery and can be put to other applications in the medical image analysis field. In future, we are planning to do a organ or tissue specific rendering. Techniques such as \cite{Shrivastava:2014} can be combined with Antara to let the doctors perform a virtual test surgery.



\end{document}